\documentclass[prl,aps,twocolumn,showpacs]{revtex4}
\usepackage{epsfig}
\newcommand{\be}{\begin{equation}}
\newcommand{\ee}{\end{equation}}
\newcommand{\bea}{\begin{eqnarray}}
\newcommand{\eea}{\end{eqnarray}}

\newcommand{\p}{\partial}

\newcommand{\re}{\mbox{e}}

\def\nn{\nonumber\\}

\def\op{{\omega^\prime}}
\def\qp{{q^\prime}}
\def\r#1{({\ref{#1}})}
\def\ep{\epsilon^\prime}
\begin{document}
\title{Finite temperature spectral function of Mott insulators and
CDW States}

\author{ Fabian H.L. Essler and  Alexei M. Tsvelik}
\affiliation{Department of  Physics, Brookhaven National Laboratory,
Upton, NY 11973-5000, USA}

\begin{abstract}
We calculate the low temperature spectral function of one-dimensional
incommensurate charge density wave (CDW) states and half-filled Mott
insulators (MI). At $T=0$ there are two dispersing features associated
with the spin and charge degrees of freedom respectively. We show that
already at very low  temperatures (compared to the gap) one of these
features gets severely damped. We comment on implications of this
result for photoemission experiments. 
\end{abstract}
\pacs{71.10.Pm, 72.80.Sk}
\maketitle

The determination of the finite temperature single-electron Green's
function in strongly interacting one-dimensional (1D) electron
systems with a spectral gap is a problem of significant interest as
the spectral function $A(\omega,q)$ is expected to exhibit the
celebrated property of spin-charge separation. Rather than a single
coherent quasiparticle peak, one expects to see 
two broad features in $A(\omega,q)$, which are associated with the 
independent dispersion of spin and charge collective modes. This
phenomenon also occurs in the case of a 1D metallic 
Luttinger liquid state. The great advantage of the insulating
state is that due to the presence of a gap it is much more robust
against the effects of 3D couplings, temperature or impurities. Hence
the chances to observe spin-charge separation by means of Angle Resolved
Photoemission Spectroscopy (ARPES) are higher in Mott or CDW
insulators than in metallic systems. Unfortunately it is much more
difficult to determine dynamical correlation functions in these
states. The very existence of a gap prohibits the application of
methods based on conformal field theory such as the Luttinger liquid
approach. A further complication is that the gap is dynamically
generated, which invalidates mean-field approximations to the
problem. In light of these difficulties, 1/N-expansions have been
a method of choice, where $N$ is a large parameter introduced by
enlarging the spin rotational symmetry of the Hamiltonian from $SU(2)$
to $SU(N)$. \\
In this letter we derive asymptotically exact expressions for the spectral 
function and the tunneling density of states for incommensurate CDW states 
and half-filled MI in the field theory limit. The reason to treat both
cases at once is that the results for the MI can be obtained from those
for the CDW state. At $T=0$ we perform our calculations for general
$N$, which allows us to see how the large-$N$ result progressively
loses its accuracy as $N$ decreases and for $N=2$ no longer
reproduces even qualitative features of the solution. We demonstrate
that Luttinger's theorem continues to hold despite the absence of a
Fermi surface. Finally we consider the effects of temperature in the
physically relevant case $N=2$ by means of a systematic expansion in
$\exp(-\Delta/T)$, where $\Delta$ is the spectral gap. 
There has been much previous work on determining the $T=0$
spectral function of 1D MI and CDW states. In Ref \cite{voit} a
conjecture for $A(\omega,q)$ was put forward, which agrees with the
result we obtain in the field theory limit by means of an exact,
systematic method. Expressions for the spectral function have also
been obtained in the limit where the single-particle gap is much
larger than the bandwodth \cite{largeU}. This regime is complementary
to the case we address here. Finally there also have been extensive
numerical studies on $t$-$J$ and Hubbard models e.g. Refs
\cite{numerical,arpes}.\\ 
{\sl Incommensurate CDW state:}
As a starting point for a microscopic description of the CDW state one
may choose a model of noninteracting electrons at some incommensurate band
filling, coupled to 1D phonons. Examples are the Su-Schrieffer-Heeger
\cite{ssh} and Holstein \cite{holstein} Hamiltonians. The electronic
low-energy degrees of freedom have momenta close to the Fermi momenta
$\pm k_F$. In the low-energy sector the electron operators can
therefore be represented as 
\be
c_{n,\sigma}=\sqrt{a_0}\left[e^{ik_Fx}\ R_\sigma(x)+
e^{-ik_Fx}\ L_\sigma(x)\right]\ ,
\ee
where $a_0$ is the lattice spacing, $x=na_0$ and $R$ and $L$ are
slowly varying Fermi fields. In three spatial dimensions the presence
of an electronic spectral gap would automatically imply the formation
of an anomalous average 
$\langle R^\dagger_{\sigma}(x)L_\sigma(x)\rangle\neq 0.$
In one dimension the average is not formed even at $T = 0$; instead
correlation functions of the operator $R^\dagger_\sigma L_\sigma$
decay in a power-law fashion.
In the limit of infinite phonon frequency $\omega_0$ the
phonons can be integrated out without inducing retardation effects in
the resulting effective electron Hamiltonian \cite{fradkin}. The case of
large but finite $\omega_0$ can be treated similarly, as long as one
is only interested in low energies $\omega\ll\omega_0$
\cite{voitphonon}. In this regime the electronic model obtained by
integrating over the phonon degrees of freedom is described by the
following universal Hamiltonian
\bea
{\cal H} &=& {\cal H}_{c} + {\cal H}_s\ ,\label{thirring}\\
{\cal H}_c &=& \frac{v_c}{16\pi}[K_c^{-1}(\p_x\Phi_c)^2 +K_c(\p_x\Theta_c)^2]\ ,\\ 
{\cal H}_s &=& \frac{2\pi v_s}{3}[:J^aJ^a: + :\bar J^a\bar J^a:] +
v_sgJ^a\bar J^a\ .
\label{curr}
\eea
Here $\Phi_c$ is a canonical Bose field, $\Theta_c$ is its dual
field and $J^a=L^\dagger_\sigma \tau^a_{\sigma\sigma^\prime}
L_{\sigma^\prime}$, $\bar J^a=R^\dagger_\sigma \tau^a_{\sigma\sigma^\prime}
R_{\sigma^\prime}$ are current operators satisfying the level-1
SU(2) Kac-Moody algebra. The parameters $v_{c,s}$ (charge and spin
velocities), $\Delta$ (spin gap) and $K_c$ (Luttinger liquid
parameter) depend on the details of the underlying microscopic lattice
model and in particular on $\omega_0$. The parameter $K_c$ controls
the scaling dimensions in the charge sector.\\
{\sl Half-filled MI:}
The Hamiltonian \r{thirring}-\r{curr} is identical to the low-energy
theory for a half-filled MI like the Hubbard model,
provided we interchange charge and spin sectors $c\leftrightarrow s$ and
then set $K_s=1$, $k_F=\pi/2$. Below we present results for the more
general CDW case with the understanding that the corresponding results
for the MI are obtained by the aforementioned mapping.\\
{\sl Zero Temperature:}
The model \r{thirring}-\r{curr} is generalized to SU(N) by replacing the
currents in \r{curr} by their SU(N) analogs. The spectrum in the charge
sector is gapless
$E_c(k) = v_c|k|$,
whereas the spectrum in the spin sector consists of scattering states
of electrically neutral, massive solitons with dispersion
$E_s(k) = \sqrt{\Delta^2 + (v_sk)^2}$ and their bound states for
$N>2$; the presence of bound states does not affect the results
presented here. For small $g\ll 1$ we have  $\Delta =
Dg^{1/N}\re^{-2\pi/Ng}$, where $D \sim \omega_0$ is the ultraviolet
cut-off. The electron operators can be expressed as products of
(vertex) operators acting in the charge and spin sectors respectively 
\bea
\langle L_\sigma(\tau,x)\ L^\dagger_\sigma(0)\rangle&=&
\prod_{\alpha=c,s}\langle{\cal O}_\alpha(\tau,x)\
\ {\cal O}^\dagger_\alpha(0)\rangle .
\eea
Using the results of \cite{gf2,LukZam01} we obtain the following
result for the asymptotics of the single-particle Green's function at
zero temperature
\bea
G(\tau,x) &=& e^{-i k_Fx}G_R(\tau,x) + e^{i k_Fx}G_L(\tau,x)\ ,\\
G_L(\tau,x) &\simeq&  \frac{Z_N\Delta}{\sqrt{\pi v_sv_c}2\pi}
\left[\frac{2v_c/\Delta}{(v_c\tau + ix)}\right]^{\frac{1}{N}}
\left[\frac{(2v_c/\Delta)^2}{x^2 +v_c^2\tau^2}\right]^{\frac{\theta}{2}}\nn
&\times&
\left(\frac{v_s\tau - ix}{v_s\tau + ix}
\right)^{\frac{N-1}{2N}}
K_{1 - \frac{1}{N}}(\Delta r)\ ,
\label{T0}
\eea
where $r=\sqrt{\tau^2 + x^2v_s^{-2}}$, $Z_N$ is a nonuniversal
constant and $\theta = \frac{1}{2N}\left[K_c+\frac{1}{K_c}-2\right]$.
We note that $G_R(\tau,x) = G_L(\tau,- x)$.
Equation \r{T0} is derived by taking into account processes with
emission of an arbitrary number of charge excitations and {\it only
one} massive spin excitation. The corrections to \r{T0} are of order
${\cal O}(e^{-3\Delta r})$ for $N=2$. However, due to the fact that matrix
elements corresponding to the emission of multi-soliton states are
numerically small, the accuracy achieved by this approximation is
very good. The smallness of matrix elements also allows us to
generalize our calculations to finite $T$. In the case of equal
velocities $v_s=v_c=v$ the retarded Green's function at $T=0$ for 
right moving fermions is 
\bea
&&G_R(\omega, k_F+q) = \frac{Z_N}{\pi^{1/2}}\Gamma\left(1 - \frac{\theta}{2}\right)
\Gamma\left(2-\frac{1}{N}-\frac{\theta}{2}\right)\nn
&&\times\frac{\omega +  vq}{\Delta^2}
F\left(1 - \frac{\theta}{2},2 -\frac{1}{N}-\frac{\theta}{2},2;
\frac{s^2}{\Delta^2}\right),
\label{GrN} 
\eea
where $F$ is a hypergeometric function and $s^2 = \omega^2 -
(vq)^2$. As might be expected, the gap in the spectral function is
'clean', i.e. $A_R(\omega,k_F+q)=-{\rm Im}G_R(\omega,k_F+q)/\pi$
vanishes for $|\omega|\leq\sqrt{\Delta^2+v^2q^2}$. 
The zero
temperature tunneling density of states is
\bea
\rho(\omega)&\simeq& A_N
\int_0^{{\rm acosh}(\omega/\Delta)}\!\!\!
dx\frac{\cosh(x[1-1/N])}
{(\omega/\Delta-\cosh x)^{1 - \theta -1/N}}\ ,
\label{rhoN}
\eea
where $A_N=\frac{Z_N\ 2^{\frac{1}{N}+\theta}}{\sqrt{\pi v_sv_c}\Gamma(\theta+1/N)}$
We see that at $T=0$ $\rho(\omega)$ vanishes inside the gap. The
singularity just above the threshold ($0<(\omega/\Delta)-1\ll 1$) is
\bea
\rho(\omega) &\simeq& \frac{A_NB\left(\theta+\frac{1}{N},
\frac{1}{2}\right)}{\sqrt{2}}\ 
(\omega/\Delta-1)^{\theta+\frac{1}{N} - \frac{1}{2}}\ .
\label{power}
\eea
We note that for the physical case $N=2$ the ``asymptotic'' result
\r{power} is actually equal to \r{rhoN}. According to (\ref{power})
the behavior of $\rho(\omega)$ above the gap is determined by the
scaling exponent $1/2 - 1/N + \theta$. The non-universal part $\theta$
is small for large phonon frequencies. On the other hand, the
remaining part is determined only by the Lorentz spin of the spinon
creation operator, and therefore is universal. Equations \r{rhoN}-\r{power}
show that a $1/N$-expansion fails completely in the physically
relevant case $N=2$, where the tunneling density of states experiences
a qualitative change and becomes nonsingular at the threshold.
As we have alluded to earlier, the $N\to\infty$ limit agrees with the
mean-field results of Ref. \cite{LRA73}.\\
{\sl Luttinger's Theorem:}
The Green's function \r{GrN} has branch cuts, but no poles. In
particular there are no poles at zero frequency and thus no Fermi
surface. Nevertheless Luttinger's theorem as stated in Ref. \cite{AGD}
is fulfilled since the logarithm of the Green's function $\ln G(\omega
=0, k_F+p)$ is still singular at the non-interacting Fermi surface,
because the Green's function \r{GrN} has {\sl zeroes}.
This property follows from (i) the fact that the charge sector is
gapless, which implies that $\langle R_\sigma(\tau,x) L^\dagger_\sigma
(0)\rangle=0$ and (ii) Lorentz invariance of the low-energy effective
theory, which implies that
\bea
\langle \Psi_\sigma(\tau,x) \Psi_\sigma^\dagger
(0)\rangle&\sim&\exp(\pm i\phi){\cal R}(r);\ \Psi=R,L.
\label{rr}
\eea
Here $r$ and $\phi$ are polar coordinates and ${\cal R}$ denotes the 
radial part of the correlation function. As we
are dealing with an insulating state we have ${\cal R}(r)\propto
\exp(-\Delta r)$ at large distances and hence $\int dr\
{\cal R}(r)r$ is finite. Thus
\bea
G_{R,L}(0,0)=\int_{-\pi}^\pi d\phi\exp(\pm i\phi)\int dr\ {\cal
R}(r)r=0. 
\eea
For a metallic state the $r$ integral would diverge and the Green's
function would have a singularity rather than a zero. \\
{\sl Finite Temperature:}
Let us now turn to the general case $v_c\neq v_s$ and finite
temperatures. We will restrict the discussion to $N =2$. The
spectral function is conveniently expressed as a convolution 
\bea
&&A_R(\omega,k_F+q)=\frac{1}{(2\pi)^3}\int_{-\infty}^\infty
d\omega^\prime dq^\prime\ \tilde{g}_s(\op,\qp)\nn
&&\times\left[\tilde{g}_c(\omega-\op,q-\qp)
+\tilde{g}_c(-\omega-\op,-q-\qp)\right],
\label{al1}
\eea
where $\tilde{g}_{c,s}(\omega,q)$ are the Fourier transforms of the
finite temperature correlators 
$\langle {\cal O}_{c,s}(x,t)\ {\cal O}^\dagger_{c,s}(0,0)\rangle_T$.
In the charge sector we can use the standard conformal mapping to
obtain (see e.g. \cite{orgad})
\bea
\tilde{g}_c(\omega,q)&=&C(T)\
f\left(\frac{\omega+v_cq}{2\pi T},\frac{1+\theta}{2}\right)
f\left(\frac{\omega-v_cq}{2\pi T},\frac{\theta}{2}\right),\nn
f(\alpha,\gamma)&=&{\rm Re}\left[(-2i)^\gamma
B\left(\frac{\gamma-i\alpha}{2},1-\gamma\right) 
\right],
\label{gc}
\eea
where $C(T)=(v_c/2\pi^3)^{1/2}(2\pi/\Delta)^\theta\ T^{\theta-3/2}$
and $B(x,y)$ is the Euler Beta function.
The correlation function in the spin sector can be determined by using
exact results for the SU(2) Thirring model. We invoke a spectral
respresentation in terms of scattering states of solitons and
antisolitons, constructed by means of the Zamolodchikov-Faddeev
algebra (see e.g. p.7 of Ref.\onlinecite{smirnov}). We introduce an
index $\epsilon=\pm$ for solitons and antisolitons respectively and
parametrize energy and momentum by a rapidity variable $\theta$ in the
usual way as $E(\theta)=\Delta\cosh\theta$,
$P(\theta)=(\Delta/v_s)\sinh\theta$. In a basis of scattering states
$|\theta_n\ldots\theta_1\rangle_{\epsilon_n\ldots\epsilon_1}$ of
solitons and antisolitons with rapidities $\theta_j$ the following
spectral representation for thermal two-point functions holds
\begin{widetext}
\bea
\langle{\cal O}^\dagger(\omega,q)\
{\cal O}(-\omega,-q)\rangle_T&=&\sum_{n,\{\epsilon_j\}}\frac{1}{n!}
\int\prod_{j=1}^n\frac{d\theta_j}{2\pi}
\sum_{m,\{\ep_k\}}\frac{1}{m!}
\int\prod_{k=1}^m\frac{d\theta_k}{2\pi}
\Bigl|{^{\epsilon_1\ldots\epsilon_n}\langle  \theta_1\ldots\theta_n|
{\cal  O}(0,0)|\theta^\prime_m\ldots\theta^\prime_1
\rangle_{\ep_m\ldots\ep_1}}\Bigr|^2\nn
&\times&
e^{-{\sum_{j=1}^nE(\theta_j)/T}}\
(2\pi)^2\delta\bigl(\omega-\sum_{j=1}^nE(\theta_j)+\sum_{k=1}^mE(\theta^\prime_k)\bigr)
\delta\bigl(q-\sum_{j=1}^nP(\theta_j)+\sum_{k=1}^mP(\theta^\prime_k)\bigr).
\label{specrep}
\eea
\end{widetext}
As was shown recently, the representation \r{specrep} is suitable for
carrying out a low-temperature expansion ($T\alt \Delta$) of
correlation functions \cite{mussardo,konik}. Taking into account the
two leading terms $n=0$, $m=1$ and $n=1$, $m=0$ in \r{specrep} and
combining this result with \r{gc} and \r{al1} we obtain
\bea
&&A_{R}(\omega,k_F+q)\approx{\cal A}\int_{-\infty}^\infty\!\! dz\ e^{z/2}
\biggl[\tilde{g}_c(\omega-c(z),q-s(z))\nn
&&+\ e^{-c(z)/T}\tilde{g}_c(\omega+c(z),q+s(z)) +
\left\{\matrix{\omega\to -\omega\cr q\to -q}\right\}\biggr],
\eea
where ${\cal A}=\sqrt{\frac{\pi\Delta}{v_s}}\frac{Z_2}{(2\pi)^3}$,
$c(z)=\Delta\cosh z$ and $s(z)=(\Delta/v_s)\sinh z$.
Let us now see how changing $\theta$ and $T$ affects
$A_R(\omega,q)$. We constrain our discussion to the cases of zero 
temperature and varying $\theta$ and of finite $T$ and fixed
$\theta$. 
\begin{figure}[ht]
\begin{center}
\epsfxsize=0.35\textwidth
\epsfbox{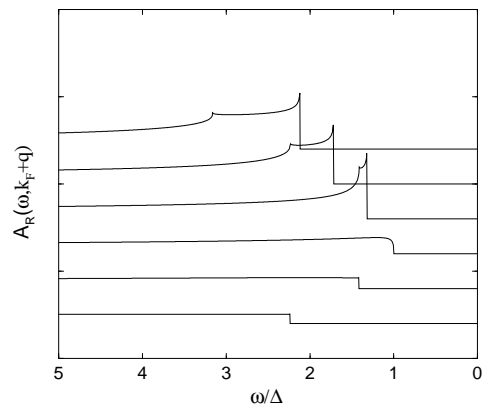}
\end{center}
\caption{\label{fig:arr}
Spectral function $A_{R}(\omega,k_F+q)$ at $T=0$, $\theta=0.8$,
$v_c=0.4 v_s$ for $v_sq/\Delta=-2,-1,0,1,2$ (from bottom to top). The
curves for different values of $q$ have been offset. 
}
\end{figure}
As a function of $\theta$, $A_R(\omega,k_F+q)$ displays a strongly varying
behaviour. For small $\theta$, corresponding to a high phonon
frequency, the spectral function is very similar to the one of the
half-filled MI (see the dotted curves in Fig.2): there
are two sharp, dispersing features associated with the spin and charge
degrees of freedom respectively. For smaller phonon frequencies and
concomitantly larger $\theta$ these features become less prominent
until they eventually disappear altogether. We plot $A_R(\omega,k_F+q)$ at
$T=0$  for the intermediate value $\theta=0.8$, $v_c=0.4\ v_s$ and
several values of $q$ in Fig.\ref{fig:arr}. We see that the peak
associated with $v_s$ is already quite weak. 
The effects of a small, finite temperature $T=0.05\Delta$ are shown
in Fig.\ref{fig:arr_T1} for a half-filled MI (we now switch spin and
charge sectors as discussed above and set $\theta=0$, $k_F=\pi/2$).
Compared to the $T=0$ result we find that the holon peak is significantly
damped although $T$ is still quite small compared to the low energy
scale $\Delta$. The physical reason for this is very simple: in the MI
only the charge sector is protected by the gap, whereas the gapless
spin sector is significantly affected by $T$. This leads to a damping
of the peak associated with the charge degrees of freedom, whereas the
spinon peak stays rather sharp. 
\begin{figure}[ht]
\begin{center}
\epsfxsize=0.45\textwidth
\epsfbox{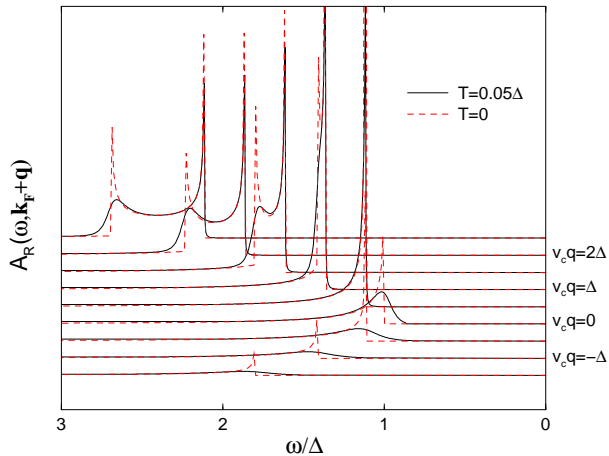}
\end{center}
\caption{\label{fig:arr_T1}
Spectral function for a half-filled MI ($\theta=0$, $v_s=0.5 v_c$),
$T=0$ (dotted) and $T=0.05\Delta$ (solid) $v_cq/\Delta=-1.5,-1,\ldots
2.5$ (from bottom to top). The curves for different values of $q$ have
been offset.}
\end{figure}

We note that ARPES measurements on 1D cuprate Mott insulators
\cite{arpes} are taken at room temperature in order to avoid charging
of the sample. The temperature used in Fig.\ref{fig:arr_T1} is chosen
to reflect the ratio $T/\Delta$ in the experiments on ${\rm
  Sr_2CuO_3}$. Although the theory presented here does not describe
the 1D cuprate MI quantitatively ($U/t$ is too large in these
compounds), we believe that our result gives a strong indication that
temperature effects are important and at least partially account for the
fact that the holon feature is barely visible in the data. 
The low-$T$ behavior of the tunneling density of states can be
analyzed analogously. We find
\bea
\rho(\omega)&=&\int_{-\infty}^\infty \frac{dx}{(2\pi)^3}\
{\rho}_s(x){\rho}_c(\omega-x)+\omega\to-\omega,
\label{rhoT2}
\eea
where 
\bea
&&{\rho}_c(x)=D{\rm Re}\biggl[ 
(-2i)^{\theta+\frac{1}{2}} 
B\left(\frac{1+2\theta}{4}-i\frac{\omega}{2\pi T},\frac{1}{2}-\theta
\right)\biggr],\nn
&&{\rho}_s(x)\simeq Z_2\sqrt{\frac{\pi}{v_s}}\biggl[
\theta_H(\omega-\Delta)+e^{-|\omega|/T}\ \theta_H(-\omega-\Delta)
\biggr]\nn
&&\times\left[
\frac{\sqrt{|\omega|-\sqrt{\omega^2-\Delta^2}}}
{\sqrt{\omega^2-\Delta^2}}
+\frac{\sqrt{|\omega|+\sqrt{\omega^2-\Delta^2}}}
{\sqrt{\omega^2-\Delta^2}}\right],
\label{rhoT}
\eea
where $D=(2\pi/\Delta)^\theta(8\pi/v_c)^{1/2} T^{\theta-1/2}$.
In the regime $T\ll\omega\ll\Delta$ we may use
contour techniques to extract the leading contribution to \r{rhoT2}  
\bea
\rho(\omega)\approx \frac{Z_2
2^\theta}{\Gamma(\frac{1}{2}+\theta)\sqrt{v_cv_s}}
\sqrt{\frac{T}{\Delta-\omega}}\ e^{-(\Delta-\omega)/T}.
\eea
This shows that at low temperatures only an exponentially small
fraction of spectral weight gets transferred into the gap.
In summary, we have calculated the low temperature spectral function
for incommensurate CDW states and half-filled Mott insulators in the
field theory limit. Luttinger's theorem is shown to hold
despite the absence of a Fermi surface. We have studied the effects of 
temperature on the two dispersing features associated with spin and
charge degrees of freedom. We demonstrated that already a small
temperature essentially wipes out the holon peak in the half-filled
Mott insulator.
We thank A. Chubukov, A. Finkelstein, P.D. Johnson and T. Valla for
important discussions. This work was supported by the DOE under
contract number DE-AC02-98 CH 10886.

\end{document}